\documentclass[aps,prc,reprint,superscriptaddress]{revtex4-1}
\usepackage[pdftex]{graphicx}
\usepackage[]{amsbsy}
\usepackage[]{amsmath}

\begin{document}


\title{Study of $\boldsymbol{(\alpha,p)}$ and $\boldsymbol{(\alpha,n)}$ reactions with a Multi-Sampling Ionization Chamber}


\author{M. L. Avila}
\email[]{mavila@anl.gov}
\affiliation{Physics Division, Argonne National Laboratory, Argonne IL 60439, USA}

\author{K. E.  Rehm}
\affiliation{Physics Division, Argonne National Laboratory, Argonne IL 60439, USA}

\author{S. Almaraz-Calderon}
\affiliation{Department of Physics, Florida State University, Tallahassee, FL 32306, USA}

\author{A. D. Ayangeakaa}
\affiliation{Physics Division, Argonne National Laboratory, Argonne IL 60439, USA}

\author{C. Dickerson}
\affiliation{Physics Division, Argonne National Laboratory, Argonne IL 60439, USA}

\author{C. R. Hoffman}
\affiliation{Physics Division, Argonne National Laboratory, Argonne IL 60439, USA}
 
\author{C. L. Jiang}
\affiliation{Physics Division, Argonne National Laboratory, Argonne IL 60439, USA}

\author{B. P. Kay}
\affiliation{Physics Division, Argonne National Laboratory, Argonne IL 60439, USA}

\author{J. Lai}
\affiliation{Department of Physics and Astronomy, Louisiana State University, Baton Rouge, LA, 70803, USA }

\author{O. Nusair}
\affiliation{Physics Division, Argonne National Laboratory, Argonne IL 60439, USA}

\author{R. C. Pardo}
\affiliation{Physics Division, Argonne National Laboratory, Argonne IL 60439, USA}

\author{D. Santiago-Gonzalez}
\affiliation{Department of Physics and Astronomy, Louisiana State University, Baton Rouge, LA, 70803, USA }
\affiliation{Physics Division, Argonne National Laboratory, Argonne IL 60439, USA}

\author{R. Talwar}
\affiliation{Physics Division, Argonne National Laboratory, Argonne IL 60439, USA}

\author{C. Ugalde}
\affiliation{Physics Division, Argonne National Laboratory, Argonne IL 60439, USA}


\date{\today}

\begin{abstract}

A large number of $(\alpha,p)$ and $(\alpha,n)$ reactions are known to play a fundamental role in nuclear astrophysics. This work presents a novel technique to study these reactions with the active target system MUSIC whose segmented anode allows the investigation of a large energy range of the excitation function with a single beam energy. In order to verify the method, we performed a direct measurements of the previously measured reactions $^{17}$O$(\alpha,n)^{20}$Ne, $^{23}$Na$(\alpha,p)^{26}$Mg, and $^{23}$Na$(\alpha,n)^{26}$Al. These reactions were investigated in inverse kinematics using $^{4}$He gas in the detector to study the excitation function in the range of about 2 to 6 MeV in the center of mass. We found good agreement between the cross sections of the $^{17}$O$(\alpha,n)^{20}$Ne reaction measured in this work and previous measurements. Furthermore we have successfully performed a simultaneous measurement of the $^{23}$Na$(\alpha,p)^{26}$Mg and $^{23}$Na$(\alpha,n)^{26}$Al reactions.

\end{abstract}

\pacs{}

\maketitle

\section{Introduction}

Helium, the second most abundant element in the universe, plays an important role in nuclear astrophysics through a variety of $\alpha$-particle induced reactions. For instance, the $^{13}$C$(\alpha,n)^{16}$O and $^{22}$Ne$(\alpha,n)^{25}$Mg reactions which are the main sources of neutrons in the slow neutron-capture process (s-process). Moreover, many $(\alpha,n)$ reactions have been found to be relevant for the nucleosynthesis of light nuclei in the rapid neutron-capture process (r-process) in neutrino-driven winds \cite{Pereira16}.  Other examples of  $\alpha$-induced reactions occur in the so-called $(\alpha,p)$ process \cite{Wallace81}, which is a reaction sequence of $(\alpha,p)$ and $(p,\gamma)$ reactions including the $^{14}$O$(\alpha,p)^{17}$F, $^{18}$Ne$(\alpha,p)^{21}$Na, $^{22}$Mg$(\alpha,p)^{25}$Al, $^{26}$Si$(\alpha,p)^{29}$P, $^{30}$Si$(\alpha,p)^{33}$Cl and $^{34}$Ar$(\alpha,p)^{37}$K reactions thought to occur in X-ray bursts, i.e., thermonuclear explosions on the surface of accreting neutron stars. The reaction path in the $(\alpha,p)$ process proceeds through a region of $\beta$-unstable nuclei located on the proton-rich side of the valley of stability and therefore requires experiments with radioactive ion beams done in inverse kinematics. Very few of these $(\alpha,p)$ reactions have been studied directly in the past, and so for most cases the astrophysical reaction rates of the $(\alpha,p)$ process were estimated by experiments measuring their time-inverse $(p,\alpha)$ reactions, which however, only probes the ground-state to ground-state transition. 

For systems involving stable nuclei, the experiments can use thin targets which are bombarded by high-intensity beams of $\alpha$-particles. The outgoing protons and neutrons can be detected in arrays of particle detectors, covering a large fraction of the angular distributions. After integrating the angular distribution the experiment is then repeated at a slightly different energy resulting in an excitation function covering the energy range of interest. This method can be quite time consuming due to the large number of energy changes required. An advantage is the good energy resolution that can be obtained by using thin targets, which can give information about the resonant structure in the compound system. While resonances play a role in nuclear astrophysics as well, the astrophysical reaction rates in quiescent and explosive stellar environments is calculated with energy-averaged cross sections. This information can be obtained by activation, i.e. by bombarding stacks of thicker target foils with beams of $\alpha$-particles and stopping the short-ranged radioactive reaction products in the target or in separate catcher foils. This technique works only for unstable reaction products with sufficiently long half-lives and appropriate decay properties \cite{Keedy66}. While this method can measure an excitation function in an experiment with one beam energy, it requires targets with good homogeneity and well-known thicknesses and the knowledge of the energy loss of the beams in the target material.

The goal of this paper is to describe a novel technique that can measure the angle-integrated and energy-averaged cross section covering a considerable part of the excitation function in one measurement by using an active target ionization chamber. The MUlti-Sampling Ionization Chamber (MUSIC) has a segmented anode that allows the investigation of a large part of an excitation function with a single measurement \cite{Carnelli14,Carnelli15}, thus, offering the opportunity to study the cross sections of astrophysically relevant $(\alpha,p)$ and $(\alpha,n)$ reactions. In order to verify the experimental technique we have chosen to measure the $^{17}$O$(\alpha,n)^{20}$Ne reaction in a previously studied excitation energy range. We have also remeasured the $^{23}$Na$(\alpha,p)^{26}$Mg and $^{23}$Na$(\alpha,n)^{26}$Mg reactions which are important for the production of $^{26}$Al in massive stars.

A similar detector based on a proportional counter has been used in the past for a study of the $^8$Li$(\alpha,n)^{11}$B reaction \cite{Mizoi00,Miyatake04}. At the energies used in these experiments, however, the $(\alpha,p)$ channel leading to Be particles ($Z=4$) was energetically forbidden and, thus, for this relatively light system the separation between the $^8$Li beam ($Z=3$) and the $^{11}$B reaction products ($Z=5$) was quite large and simplifies the identification of  the particles of interest. One of the goals of our study is to demonstrate that the reaction products of the $(\alpha,p)$ and $(\alpha,n)$ reactions can be separated and measured simultaneously. For this, we use the $^{23}$Na+$^4$He system in an energy range where both the $(\alpha,p)$ and $(\alpha,n)$ reaction channels are energetically allowed.

This paper is organized as follows.  In Section II we give a brief description of the MUSIC detector and discrimination of the $(\alpha,p)$ and $(\alpha,n)$ reaction products. Section III presents the comparison of  the $^{17}$O$(\alpha,n)^{20}$Ne reaction measured with MUSIC and the data obtained by traditional techniques in normal kinematics. The cross section of the $^{17}$O$(\alpha,p)^{20}$Ne reaction has a sufficiently large negative Q-value (Q $=-5.655$ MeV) so that it does not contribute in the energy range studied in the experiment. Section IV presents the study of the $^{23}$Na$(\alpha,p)^{26}$Mg and $^{23}$Na$(\alpha,n)^{26}$Al reactions and gives the results in an energy range where both reactions contribute. A summary and an outlook for future experiments is presented in Section V.

\section{ The MUlti-Sampling Ionization Chamber (MUSIC).}

\begin{figure}
\centering
\includegraphics[scale=1.2]{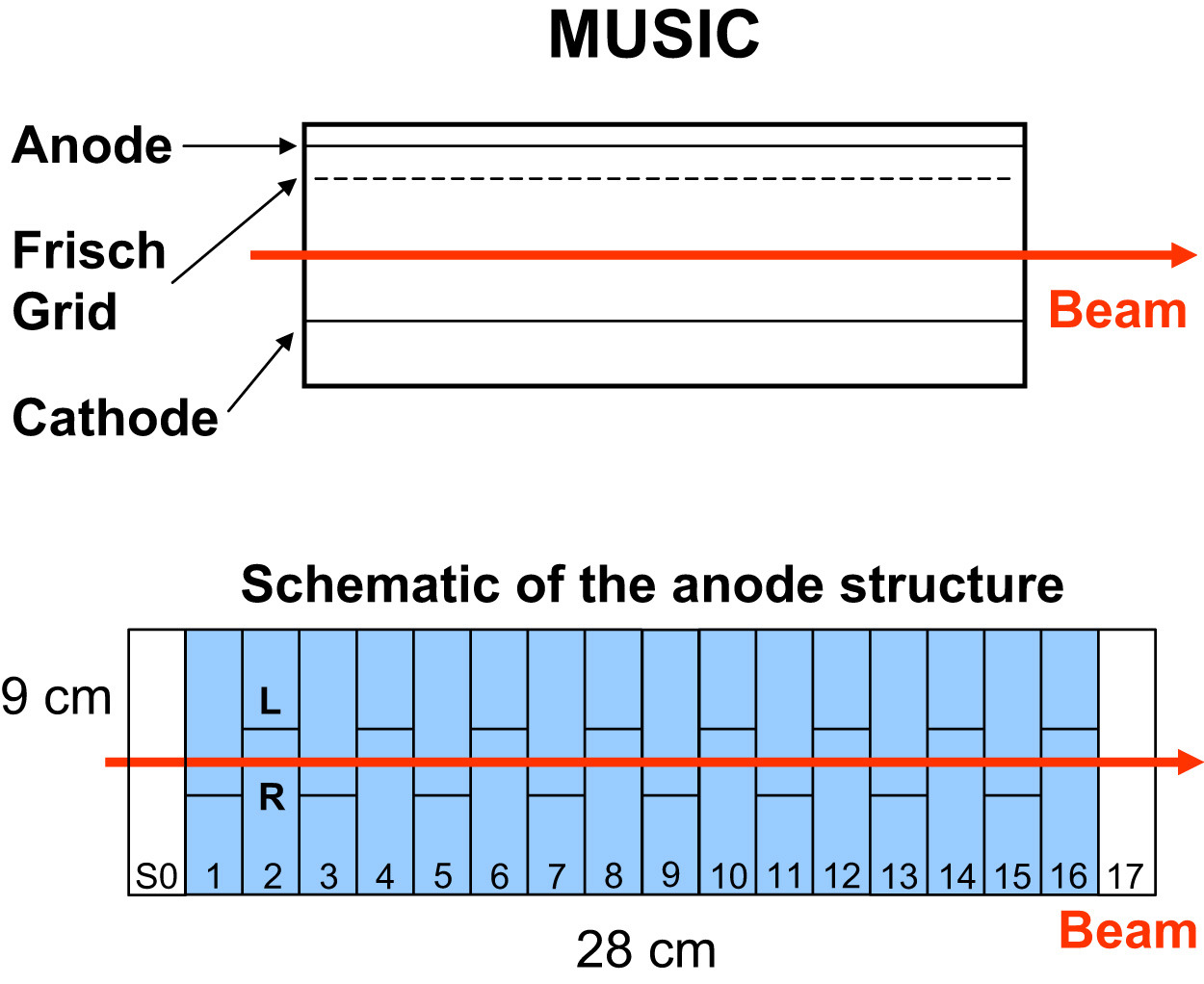} 
\caption{\label{fig:MUSIC} 
Schematic of the MUSIC detector. Upper panel: lateral view of the detector. Lower panel: anode structure, showing the 18 strips. Strips 1-16 are subdivided into non-symmetric left and right sections.}
\end{figure}

The MUlti-Sampling Ionization Chamber (MUSIC) has been previously used for measurements of fusion reactions involving radioactive nuclei \cite{Carnelli14}. We now explore its applicability to measure $(\alpha,p)$ and $(\alpha,n)$ reactions. Since a more detailed description of the detector has already been given in a separate publication \cite{Carnelli15} we will only summarize the main operation principles here and focus on the analysis of the $(\alpha,p)$ and $(\alpha,n)$ measurements. A cross section of the detector is shown in Fig. \ref{fig:MUSIC}. Electrons from the ionization of the gas molecules generated by the beam particles as they travel through the chamber volume, drift pass through the Frisch grid towards the anode which is divided into 18 strips. Strips 1-16 are subdivided in a right and left section as shown in the lower part of Fig. \ref{fig:MUSIC}. In Ref. \cite{Carnelli15} the ionization chamber was filled with CH$_4$ gas for a study of fusion reactions between $^{12}$C from the counting gas and various incident carbon ions ranging from $^{10}$C to $^{15}$C. The evaporation residues produced in these fusion reactions (O, F, Ne and Na isotopes) are at least 2 units in charge away from the incident carbon beams and therefore could easily be separated from the beam due to their considerably larger energy loss. 

\begin{figure}
\centering
\includegraphics[scale=0.45]{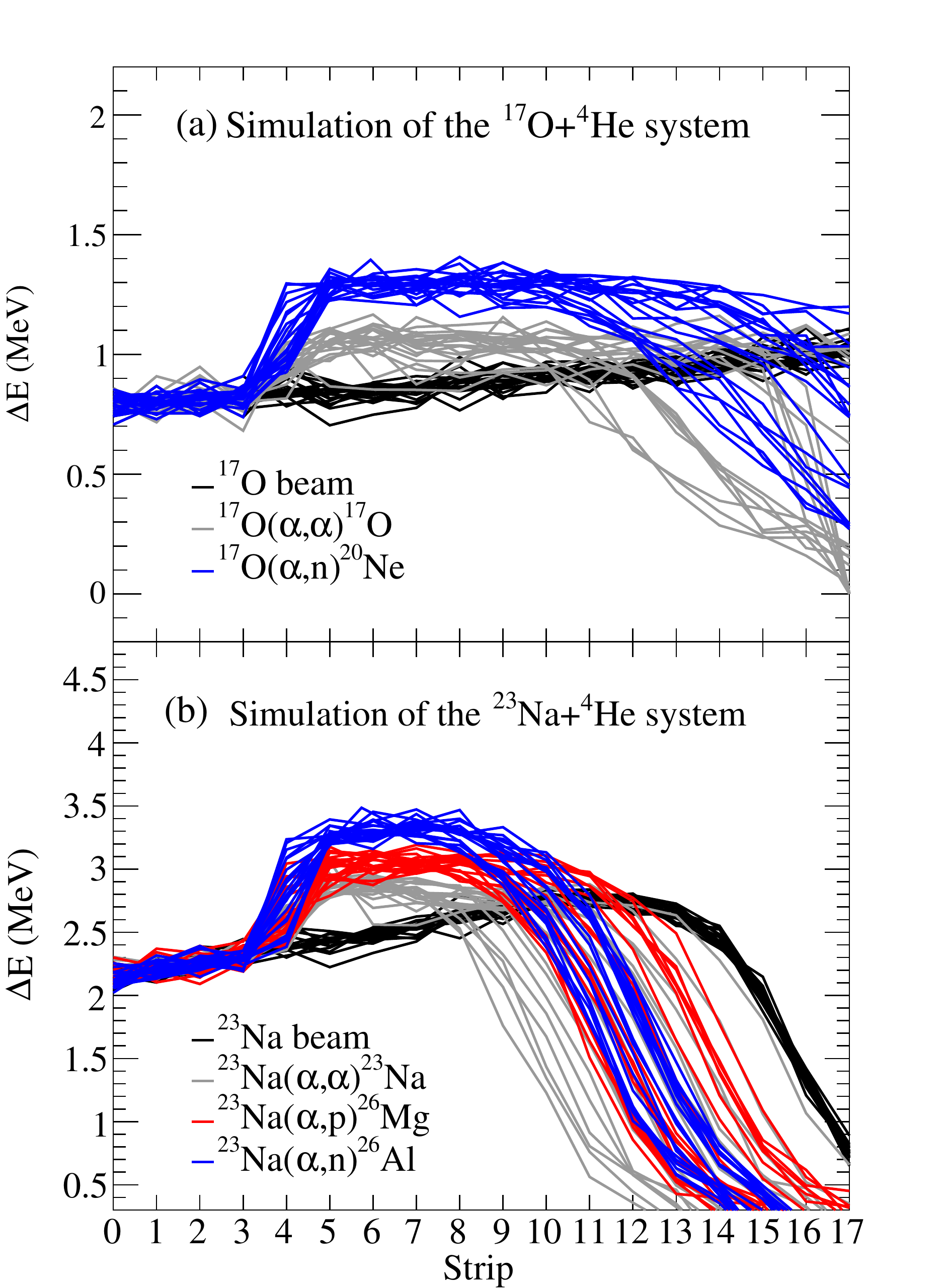} 
\caption{\label{fig:Simulation} 
Simulation of the energy-loss signals in the 18 strips of the detector MUSIC for events of the $^{17}$O$(\alpha,n)^{20}$Ne reaction (a), and the  $^{23}$Na($\alpha,p)^{26}$Mg and $^{23}$Na($\alpha,n)^{26}$Al reactions (b) occurring in strip 4. Traces of the $(\alpha,p)$ reaction are shown in red, of the $(\alpha,n)$ reaction in blue, of the $(\alpha,\alpha')$ reaction in gray while the beam traces are given in black.}
\end{figure}

For a study of the $(\alpha,p)$ and $(\alpha,n)$ reactions, the MUSIC detector was filled with helium as a counting gas at pressures ranging from about 200 to 400 Torr. Beams of $^{17}$O ($E=$ 34.8 MeV) and $^{23}$Na ($E=$ 51.5 and 57.4 MeV) from the ATLAS accelerator at Argonne National Laboratory with intensities up to 5000 particles/sec, passed through the detector.  If an incoming $^{17}$O or $^{23}$Na particle interacts with a helium nucleus through an $(\alpha,n)$ or $(\alpha,p)$ reaction, the outgoing residual nuclei $^{20}$Ne, $^{26}$Al or $^{26}$Mg are emitted at angles below 10$^{\circ}$, which is well within the angular acceptance of the ionization chamber. Compared to the fusion reactions discussed in Ref. \cite{Carnelli14}, the separation of the $(\alpha,p)$ and $(\alpha,n)$ reaction products, however, is more challenging due to their closeness in energy loss ($\Delta$E) to the incoming beam. Simulations of the energy loss of the $^{17}$O and $^{23}$Na beams passing through the 16 individual sections of the ionization chamber are shown by the black traces in Fig. \ref{fig:Simulation}. For the simulation, an isotropic distribution of the $(\alpha,p)$ and $(\alpha,n)$ reaction products is assumed. Traces of the reaction products $^{20}$Ne (for the $^{17}$O beam) and $^{26}$Al, $^{26}$Mg (for the $^{23}$Na beam) generated in strip  4 are shown in Fig. \ref{fig:Simulation} (a) and (b), respectively. Traces of the reaction products from $(\alpha,n)$ reactions are shown in blue, the $(\alpha,p)$ traces are given in red, and the $(\alpha,\alpha')$ are in gray. As can be seen from these calculations, a separation of the $(\alpha,n)$ and $(\alpha,p)$ reactions based on their different energy loss is possible. Although, the separation between the $(\alpha,p)$ and $(\alpha,\alpha')$ reactions is not as obvious in Fig. \ref{fig:Simulation} (b), these reactions can be separated using additional conditions as will be explained in section \ref{23Na}. In addition to providing angle-integrated cross sections this technique also integrates the cross section to all excited states for the simultaneously study of $(\alpha,p)$ and $(\alpha,n)$ reactions. Due to the energy loss of the incident beam in the counting gas and the segmented anode of the ionization chamber this technique provides a measurement of the energy-averaged cross sections in an energy range which is determined by the pressure used in the ionization chamber. This technique is also ``self-normalizing'' (no additional monitor detectors are needed) since the reaction products as well as the incident beam particles are measured in the detector simultaneously. 

A limitation of this method is the low beam intensity (up to 5$\times10^3$ particles/sec) that can be presently used in the ionization chamber. Thanks to the 100\% detection efficiency, however, cross sections down to the 1 mb range have been measured. Future improvements, such as the addition of faster trigger detectors before and after MUSIC or the use of a gating grid \cite{Hashimoto06}, could improve this limitation by more than an order of magnitude.

\section{The $\boldsymbol{^{17}\mathrm{O}(\alpha,n)^{20}\mathrm{Ne}}$ reaction}\label{17O}

As a first test case of MUSIC we have investigated the $^{17}$O$+^4$He system.  The $^{17}$O$(\alpha,n)^{20}$Ne reaction has been studied by \citet{Bair73} by bombarding a thin $^{17}$O target ($\sim$35 keV for 1 MeV $\alpha$ particles) with $\alpha$-particle beams in the c.m. energy range of 0.75-4.3 MeV. The outgoing neutrons from the $(\alpha,n)$ reactions were detected in a graphite sphere neutron detector \cite{Macklin57}. Due to a negative Q value of $-5.656$ MeV the $^{17}$O$(\alpha,p)^{20}$F reaction, it is energetically forbidden in this energy range. The cross sections in the energy region of 0.75-4.3 MeV ranges from about 20 $\mu$b to 190 mb and exhibits a pronounced resonant structure.
 
In our experiment we used a 34.8 MeV $^{17}$O beam delivered by the superconducting linear accelerator ATLAS. In order to reduce the beam intensity to the 5-10 kHz range, it was attenuated by a series of ‘pepper-pot’ attenuators \cite{Kubik87} and by making use of the ATLAS beam sweeper. The beam sweeper increased the pulse period of the beam from 82 ns to 41 $\mu$s, which is long compared to the drift time of the electrons in MUSIC, even for the slow drift velocities of helium. The MUSIC detector was filled with 206 Torr of $^4$He gas. The $^{17}$O beam entered the detector through a 1.45 mg/cm$^2$ thick Ti window followed by a 35 mm long dead volume of helium and reached the first strip of the anode (strip 0 in Fig. \ref{fig:Simulation} (a)) with an energy of 25.1 MeV. After passing through all 16 strips the $^{17}$O energy has decreased to 11.8 MeV. These values have been calculated by using the energy loss values of the computer code SRIM (version 2008) \cite{SRIM}. To ensure that only events with one-particle per bunch passing through the detector were selected, a condition was set on the cathode signal and on the signal of strip 0 which eliminated events with more than one particle per beam bunch. Since this experiment was done with a stable $^{17}$O beam no further conditions to identify the incoming beam were required (see e.g. Ref. \cite{Carnelli15}).

\begin{figure}
\centering
\includegraphics[scale=0.45]{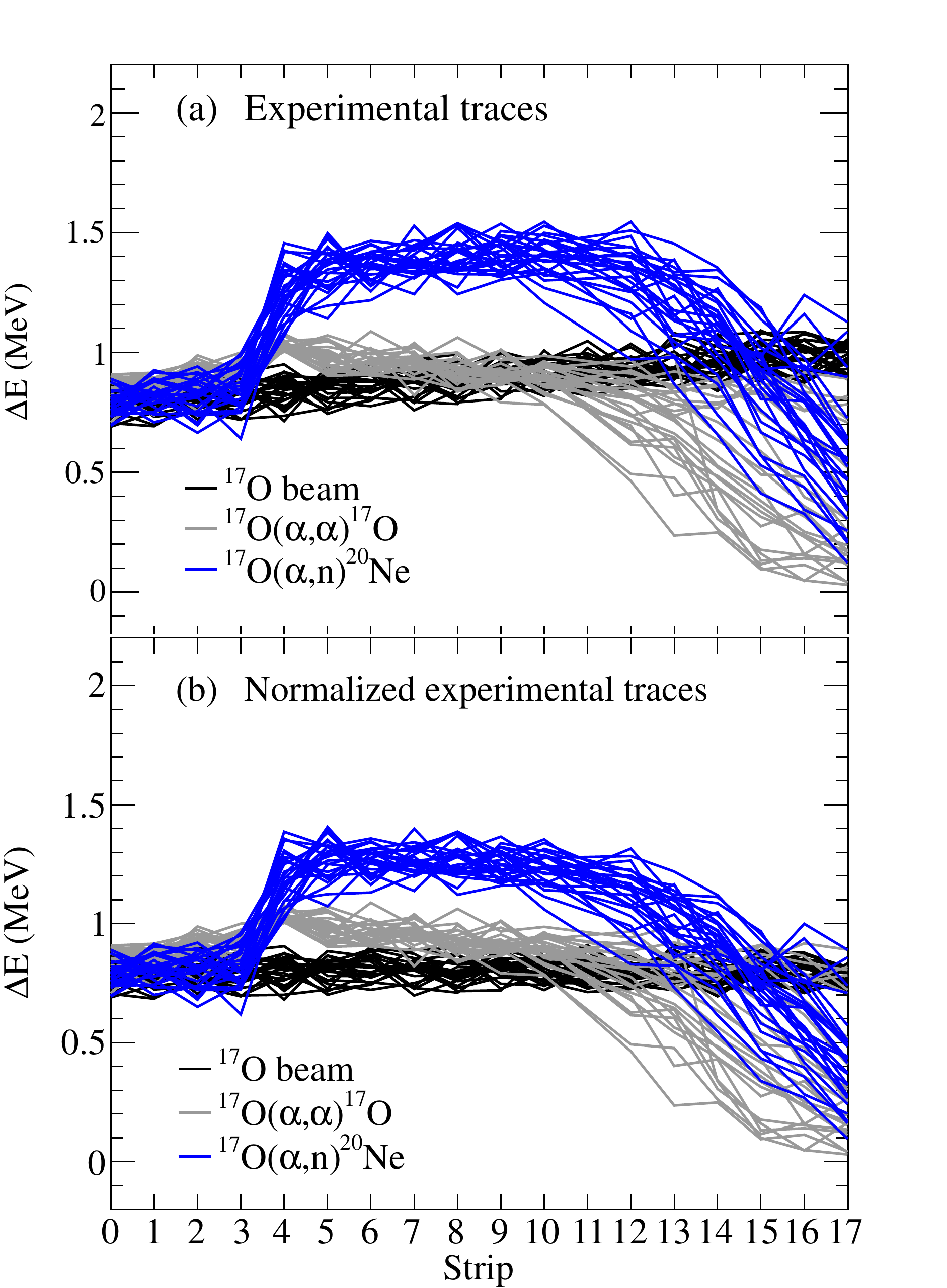} 
\caption{\label{fig:17O_strip4} 
(a) Energy-loss signals measured in the 18 strips of the MUSIC detector for events of the $^{17}$O$(\alpha,n)^{20}$Ne (blue) and the $^{17}$O$(\alpha,\alpha')^{17}$O (gray) reactions occurring in strip 4 and for the $^{17}$O beam (black). (b) Same as (a) but with the $\Delta$E values of all strips normalized to the $\Delta$E value of strip 0.}
\end{figure}

In Fig. \ref{fig:17O_strip4} (a) the energy loss signals $\Delta$E for events of the $^{17}$O$(\alpha,n)^{20}$Ne (blue) and $^{17}$O$(\alpha,\alpha')^{17}$O (gray) reactions occurring in strip 4 and of the $^{17}$O beam (black) for three runs of one hour each are shown. For a better display Fig. \ref{fig:17O_strip4} (a) only shows the first 25 events of the $(\alpha,\alpha')$ reaction. In order to simplify the analysis, the energy dependence of the $\Delta$E values has been eliminated by normalizing the beam-like signals of all anode strips to the energy loss of strip 0, as shown in Fig \ref{fig:17O_strip4} (b). It can be seen from Fig. \ref{fig:17O_strip4} (a) that the traces from the $^{20}$Ne reaction products are well separated from traces in the vicinity of the beam-like particles, and agree with the simulation shown in Fig. \ref{fig:Simulation} (a). 

\begin{figure}
\centering
\includegraphics[scale=0.45]{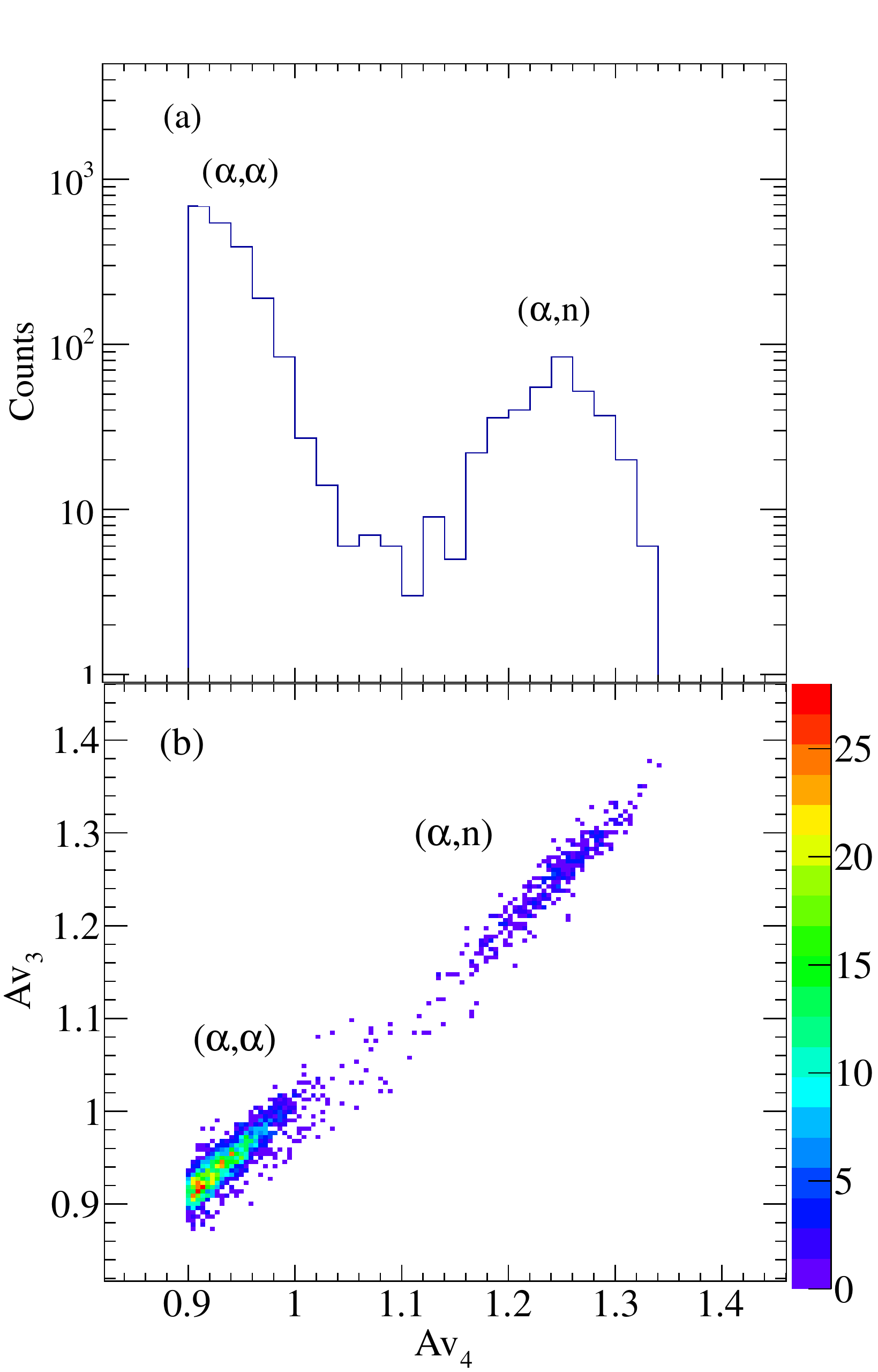} 
\caption{\label{fig:17O_Ave1_Ave2} 
(a) $\Delta$E values for events occurring in strip 4 averaged over four consecutive strips in order to improve the separation of events from the $^{17}$O$(\alpha,n)^{20}$Ne and $^{17}$O$(\alpha,\alpha')^{17}$O reactions. (b) Two-dimensional plot of the same events using averages over four (Av$_4$) and  three strips (Av$_3$), respectively.}
\end{figure}

To separate events coming from elastic scattering a multiplicity condition was used in a similar manner as explained in Ref. \cite{Carnelli14}. This eliminates most of the elastic/inelastic scattering events which have  energy loss values close to the ones from the $^{17}$O$(\alpha,n)^{20}$Ne reaction of interest. To further improve the separation between the traces from the $^{17}$O$(\alpha,n)^{20}$Ne reaction and elastic/inelastic scattering events of the $^{17}$O$(\alpha,\alpha')^{17}$O reaction, we have averaged the energy loss values of several strips following the strip where the reaction took place. This average helps to prevent an incorrect assigment of the traces to different reactions due to fluctuation of the signals if only one strip is taken into account. This average is called Av$_{n}$ with n indicating the number of strips used in calculating the average. This is shown in Fig. \ref{fig:17O_Ave1_Ave2} (a) where the average is done over 4 strips for all the events occurring in strip 4 during a 1.5 day long run. In figure \ref{fig:17O_Ave1_Ave2} (b) a two-dimensional plot of Av$_4$ vs Av$_3$ shows that events of $^{17}$O$(\alpha,n)^{20}$Ne and $^{17}$O$(\alpha,\alpha')^{17}$O can be clearly separated. The sharp cut seen in Fig. \ref{fig:17O_Ave1_Ave2} at 0.9 MeV in the $x$-axis (Av$_4$) is due to a condition applied to the data in order to discard the beam-like events.

From the number of $^{20}$Ne traces the energy-averaged cross sections of the $^{17}$O$(\alpha,n)^{20}$Ne reaction could be determined covering the energy range $E_{c.m.}= 2.6$-4.8 MeV. For the cross section calculation we calibrated the pressure readout system of the gas-handling system against a high-precision gauge from Wallace and Tiernan \cite{Wallace}. Corrections to pressure for temperature variations during the one and a half days long run were also made. Since the energy loss values used in the simulation were taken from the SRIM code, the $\Delta$E values used for the analysis were also calculated from SRIM. Changing the energy loss values to the LISE++ predictions \cite{LISE} changed the energy by about 10\% on average. The finite width of the anode strips averages the cross sections over typically 160 keV in the center-of-mass system.

\begin{figure}
\centering
\includegraphics[scale=0.29]{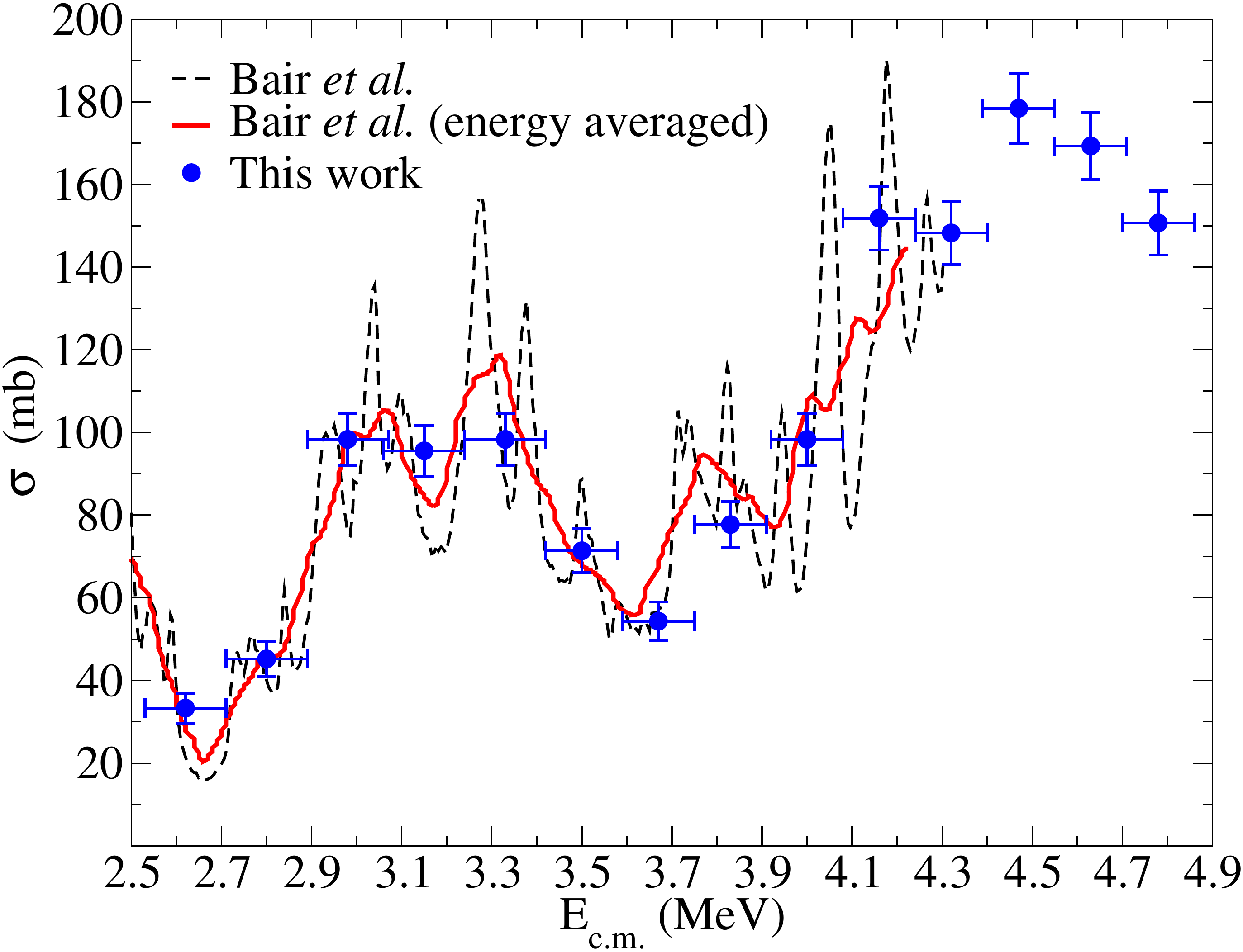} 
\caption{\label{fig:17O} 
Cross section of the $^{17}$O($\alpha,n)^{20}$Ne reaction as a function of center-of-mass energy, obtained from Ref. \cite{Bair73} (black dashed line) and its energy-averaged cross section (solid red line) compared with the MUSIC data (blue circles).}
\end{figure}

The resulting excitation function is shown in Fig. \ref{fig:17O} (blue solid points) in comparison with the cross sections from Ref. \cite{Bair73}. While the dashed black line are the data taken from the Brookhaven Data Compilation the red solid line has been averaged over the 160 keV range which is the energy width of an individual anode strip. The uncertainties in $\sigma$ are statistical and the uncertainties in energy are calculated from the energy width of each anode strip (in $E_{c.m.}$). The cross sections from MUSIC are in good agreement with the ones from the direct measurement of Ref. \cite{Bair73}, successfully proving that the MUSIC detector can be used to measure $(\alpha,n)$ reactions.

\section{The $\boldsymbol{^{23}\mathrm{Na}(\alpha,p)^{26}\mathrm{Mg}$ and $\boldsymbol{^{23}\mathrm{Na}(\alpha,n)^{26}\mathrm{Al}}}$ reactions}\label{23Na}

The $^{23}$Na($\alpha,p)^{26}$Mg  and $^{23}$Na($\alpha,n)^{26}$Al reactions are important for our understanding of the $^{26}$Al production in massive stars ($M>8M_\odot$) \cite{Illiadis11}. The $^{23}$Na($\alpha,p)^{26}$Mg reaction has recently seen increased interest due to the large cross sections reported in Ref. \cite{Almaraz14}. Although the physics behind these reactions is quite interesting, in this work we will focus on the experimental technique only. A secon level of complexity is added when both $(\alpha,p)$ and $(\alpha,n)$ channels are open. Therefore the aim of this work was to demonstrate that the MUSIC detector is able to separate and measure these reactions simultaneously. The physics case is presented in a different publication \cite{Avila16}.

\begin{figure}
\centering
\includegraphics[scale=0.45]{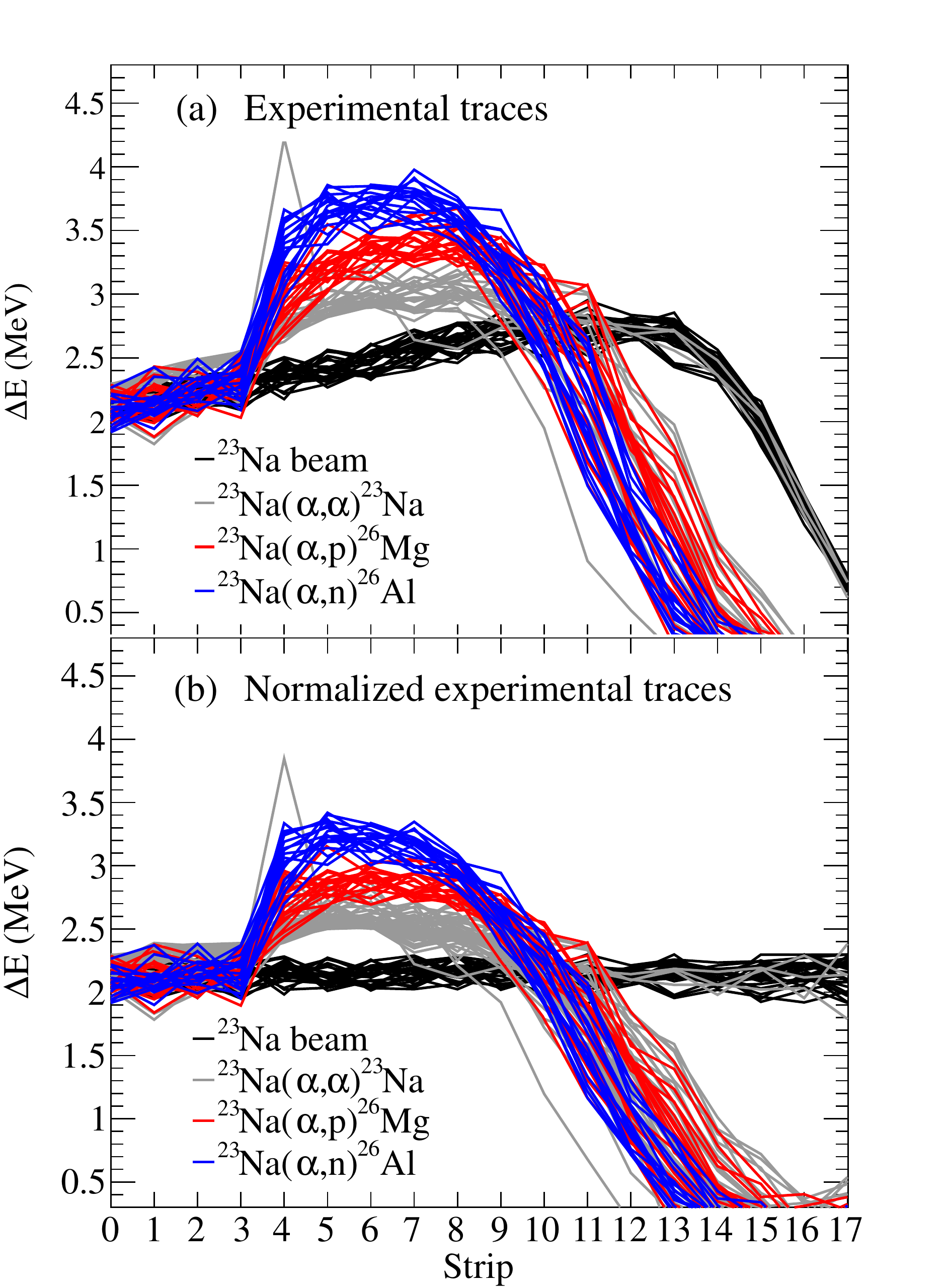} 
\caption{\label{fig:traces_23Na} 
(a) Energy-loss signals measured in the 18 strips of the MUSIC detector for events of the $^{23}$Na$(\alpha,n)^{26}$Al (blue), $^{23}$Na$(\alpha,p)^{26}$Mg (red), and $^{23}$Na$(\alpha,\alpha')^{23}$Na (gray) reactions occurring in strip 4 and the $^{23}$Na beam (black). (b) Same as (a) but using energy loss values normalized to the $\Delta$E of the beam-like particles in strip 0.}
\end{figure}

The experiment was carried out at the ATLAS accelerator at Argonne National Laboratory using a $^{23}$Na beam with energies of 51.5 and 57.4 MeV.  The beam was delivered to the MUSIC detector which was filled with 403 and 395 Torr of $^4$He gas for the lower and higher energy, respectively. The setup was identical to the one used for $^{17}$O, which was described in Sect. \ref{17O}. The energy range in the center of mass (c.m.) covered in this experiment was $E_{c.m.}=2.2$-5.8 MeV. Due to the $Q$ values of $Q=-2.967$ MeV $(\alpha,n)$ and $Q=1.820 MeV$ $(\alpha,p)$ both reaction channels are energetically allowed. Traces of the reaction products occurring e.g. in strip 4 of the MUSIC detector for a one hour run at the higher beam energy are shown in Fig. \ref{fig:traces_23Na}. Four groups of traces with increasing $\Delta$E values are visible in Fig. \ref{fig:traces_23Na} (a). They originate from the $^{23}$Na beam (black), $^{26}$Mg ions from the $(\alpha,p)$ reaction (red), $^{26}$Al ions from the $(\alpha,n)$ reaction (blue), and from elastic and inelastic scattering reactions (gray). For a better visualization only the first 25 events of the $(\alpha,\alpha')$ reaction are shown. The energy of the $^{23}$Na ions passing through strip 0 was about 39 MeV and for a pressure of 395 Torr the beam was almost stopped at strip 16, as can be seen in Fig. \ref{fig:traces_23Na} (a). The experimental traces seen in  Fig. \ref{fig:traces_23Na} (a) are in good agreement with the simulated traces shown in Fig. \ref{fig:Simulation} (b). For the analysis, the traces were again normalized to the $\Delta$E value of strip 0 as shown in Fig. \ref{fig:traces_23Na} (b).

\begin{figure}
\centering
\includegraphics[scale=0.45]{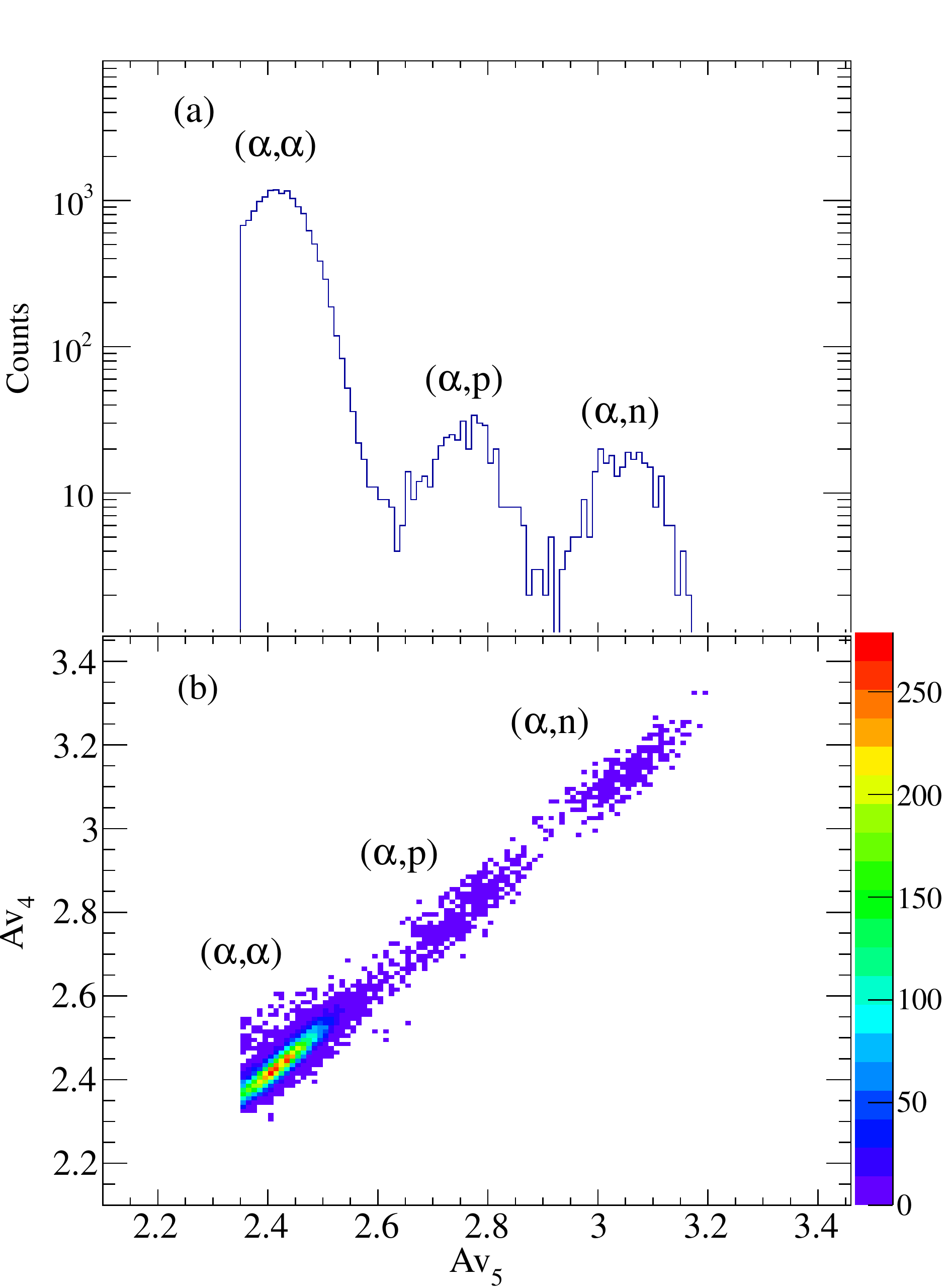} 
\caption{ \label{fig:Ave1_Ave2} 
(a) $\Delta$E values for events occurring in strip 4, averaged over five consecutive strips in order to improve the separation of events from the the $^{23}$Na$(\alpha,\alpha')^{23}$Na, the $^{23}$Na$(\alpha,p)^{26}$Mg and the $^{23}$Na$(\alpha,n)^{26}$Al reactions. (b) Two-dimensional plot of the same events using averages over five (Av$_5$) and four strips (Av$_4$), respectively. }
\end{figure}

A better separation of the three reactions can be again obtained by averaging the $\Delta$E values over a certain number of strips as mentioned for the $^{17}$O system in Sect. \ref{17O}. This is shown in Fig. \ref{fig:Ave1_Ave2} (a), where the $\Delta$E values of events occurring in strip 4 were averaged over 5 strips ($Av_5$). For $(\alpha,p)$ and $(\alpha,n)$ events occurring in the first strips the average can be  done over a larger number of strips due to the fact that the traces are longer. A good separation of the different types of events can be seen in a two-dimensional plot (see Fig. \ref{fig:Ave1_Ave2} (b)) where a five-strip average (Av$_5$) is plotted against a four-strip average (Av$_4$). In this plot a clear separation between the events coming from $(\alpha,\alpha')$ (first group) , $(\alpha,p)$ (second group) and $(\alpha,n)$ (third group) is observed. With this approach the cross sections of the $^{23}$Na$(\alpha,p)^{26}$Mg and $^{23}$Na$(\alpha,n)^{26}$Al reactions have been determined covering the energy range E$_{c.m.}\approx$2-6 MeV in the center-of-mass frame. The normalization of the cross section is performed by using the number of beam particles which are simultaneously measured in the detector.

\begin{figure}[]
\centering
\includegraphics[scale=0.3]{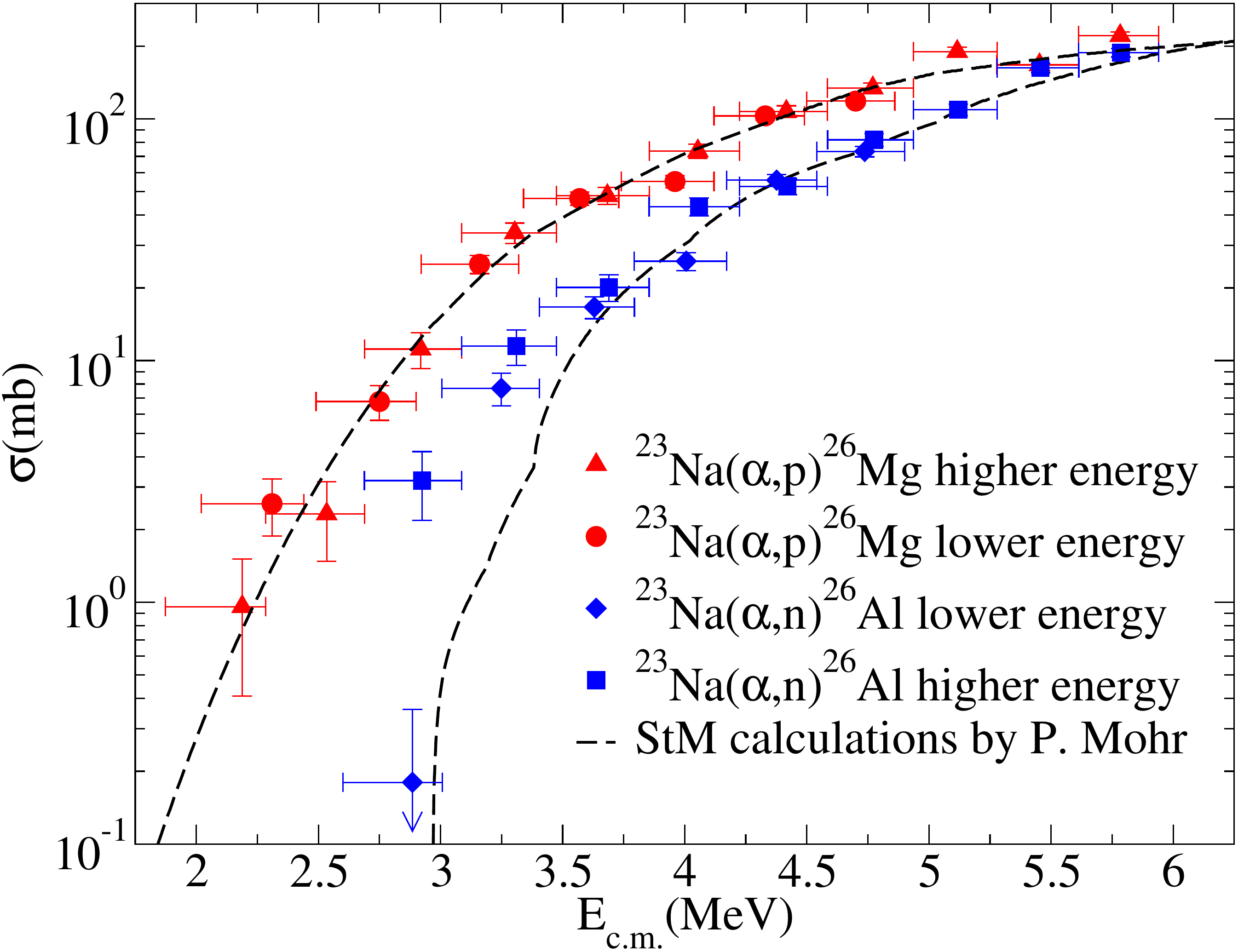} 
\caption{\label{fig:ap_an_comparison} 
Excitation functions  of the $^{23}$Na($\alpha,n)^{26}$Al reaction (blue) and the $^{23}$Na($\alpha,p)^{26}$Mg reaction (red) measured with  the MUSIC detector for the beam energies of 51.5 and 57.4 MeV labeled as low energy and high energy, respectively.}
\end{figure}

The excitation functions of the two reactions were measured simultaneously in two runs lasting about 1.5 days each for the higher and lower beam energy, respectively. The results are presented in Fig. \ref{fig:ap_an_comparison}, where the ($\alpha,p)$ data are shown by red circles for the lower beam energy and by red triangles for the higher beam energy. Similarly the ($\alpha,n)$ cross sections are shown by blue diamonds for the lower energy and blue squares for the higher beam energy, respectively. The dashed lines are the cross sections predicted for the two reactions by the statistical model from Ref. \cite{Mohr15} using the TALYS code. In our experiment the ($\alpha,p)$ and ($\alpha,n)$ were measured simultaneously with the same detector, eliminating problems arising from different detection efficiencies. The results presented in Sections \ref{17O} and \ref{23Na} give us confidence that the MUSIC detector can be used to measure $\alpha$-induced reactions.

\section{ Summary}

We have described a method to measure excitation functions of angle and excitation-energy integrated cross sections of $(\alpha,p)$ and $(\alpha,n)$ reactions which are of interest to nuclear astrophysics. Making use of a multi-sampling ionization chamber MUSIC and the advantages of inverse kinematics, the reaction products of the two reactions can be detected with 100\% detection efficiency. 

The application of the active target system MUSIC to $\alpha$-particle induced reactions was tested for the $^{17}$O$(\alpha,n)^{20}$Ne reaction. We found a good agreement between the cross section obtained with the MUSIC detector and previous measurements, demonstrating the capabilities of the detector for measuring $\alpha$-induced reactions. This was followed by a simultaneous measurement of excitation functions of the $^{23}$Na$(\alpha,p)^{26}$Mg and $^{23}$Na$(\alpha,n)^{26}$Al reactions. These measurements proved that the MUSIC detector is capable of separating ($\alpha,p$) and ($\alpha,n$) reactions based on the different energy loss of the reaction products. 

This technique can also be used for reactions with stable beams which are difficult to study with standard experimental methods. The main application of this technique, however, is with low-intensity radioactive beams, such as various $(\alpha,p)$ reactions which occur during the $(\alpha,p)$ process in X-ray bursts. 

Improvements to the technique which are presently being implemented include the addition of parallel plate avalanche detectors at the entrance and exit of MUSIC to provide a trigger signal that a reaction occurred within MUSIC. Other possible upgrades such as the addition of a gating grid and the implementation of digital electronics are under consideration.

\begin{acknowledgments}
This material is based upon work supported by the U.S. Department of Energy, Office of Science, Office of Nuclear Physics, under contract number DE-AC02-06CH11357. The authors J. L. and D. S. G. also acknowledge the support by the U.S. Department of Energy, Office of Science, Office of Nuclear Science, under Award No. DE-FG02-96ER40978. This research used resources of ANL's ATLAS facility, which is DOE Office of Science User Facility.
\end{acknowledgments}

\bibliography{myrefs}

\end{document}